\documentclass[aps,a4paper, preprint, superscriptaddress,preprintnumbers,amsmath,amssymb]{revtex4-2}
\usepackage{graphicx}
\usepackage{amsmath}
\usepackage{amsfonts}
\usepackage{amssymb}
\usepackage{url}
\usepackage{hyperref}
\usepackage{subfigure}
\usepackage{array}

\begin{abstract}
We develop an  $A_4 \times Z_4 \times Z_2$ symmetry extension of Standard Model under the minimal extended seesaw (MES) mechanism which successfully predicts neutrino masses and mixings patterns.  This model breaks $\mu-\tau$ symmetry of neutrino mass matrix and explains leptonic mixing with non-zero $\theta_{13}$. We study the phenomenological results of the keV-scale sterile neutrino as a dark matter candidate along with other phenomenologies such as neutrino oscillation observables, neutrinoless double beta decay, baryogenesis via leptogenesis, etc. Dirac CP-violating phase $\delta_{CP}$ and two Majorana phases $\alpha$ and $\beta$ are also calculated from the leptonic mixing matrix. Best-fit values of the model parameters and neutrino observables are calculated from $\chi^2$ analysis. The model predicts best-fit values of neutrino mixing angles to be $\sin^2\theta_{23}=0.555,\ \sin^2\theta_{12}=0.301$ and $\sin^2\theta_{13}=0.022$ for normal hierarchy. Significant results consistent with experimental data are also observed for effective neutrino mass $m_{\beta\beta} \sim (0.97 - 5.02)$ meV, effective electron mass $m_{\beta} \sim (0.084-0.41)$eV and sum of active neutrino masses $\sum m_{i} < 0.12$ eV. The model does not favour Inverted hierarchy at the 3$\sigma$ level with the given parameter space.
\end{abstract}

\begin{document}

\title{KeV dark matter in minimal extended seesaw model and its predictions in neutrinoless double beta decay and baryogenesis.}

\author{Mayengbam Kishan Singh}
\email{kishan@manipuruniv.ac.in}
 \affiliation{Department of Physics, Manipur University, Imphal-795003, India}                                    
  \author{S. Robertson Singh}
\email{robsoram@gmail.com}
  \author{N. Nimai Singh}
\email{nimai03@yahoo.com}
 \affiliation{Department of Physics, Manipur University, Imphal - 795003, India }   
 \affiliation{Research Institute of Science and Technology, Imphal - 795003, India}


\maketitle
\titlepage


\section{Introduction}
Inspite of enormous successes and outstanding discovery of Higg's boson, the Standard Model (SM) is incapable of explaining many problems and phenomena in the Universe. Absence of neutrino mass is one of the main shortcomings of SM. Observations of neutrino oscillation in SNO\cite{SNO,sno2002direct}, SK\cite{SK}, etc. have verified that neutrinos have non-zero mass and they can change from one flavour to another. The global analysis of the latest data from various experiments such as  solar, atmospheric, reactor and accelerator experiments, gives the best fit values of the neutrino oscillation parameters such as the two mass-squared differences ($\Delta m_{atm}^2 \sim 7.4 \times 10^{-5}$eV$^2$ and $\Delta m_{sol}^2 \sim 2.5 \times 10^{-3}$eV$^2$ ) and the three mixing angles ($\theta_{12}\sim 34^{o}, \theta_{13} \sim 7.4^{o}$ and $\theta_{23}\sim 48^{o}$)\cite{pdg2022}. Some of the burning questions in high energy physics are the absolute masses and exact nature of the neutrinos (Dirac or Majorana), Dirac CP-violating phase, baryon asymmetry of the Universe (BAU), dark matter (DM), etc.

Anomalies in the observations of LSND\cite{aguilar2001evidence} which detected an excess of electron anti-neutrino $(\overline{\nu}_e)$ in a muon anti-neutrino $(\overline{\nu}_{\mu})$ beam produced at the Los Alamos laboratory could not be explained by the three neutrino oscillation scheme. This result was further supplemented by the MiniBooNE \cite{aguilar2018significant} experiment which observed an oscillation from $(\overline{\nu}_{\mu})$ to $(\overline{\nu}_e)$ in agreement with the LSND data. These observations gave birth to the idea of the existence of additional neutrino(s) called sterile neutrino. Sterile neutrinos are singlet fermions under $SU(2)_L$ which do not have weak interactions but can mix with the active neutrinos. LSND and MiniBooNE  results could be explained by adding at least one additional sterile neutrino having mass in the eV scale to the SM in a $3+1$ framework. However, sterile neutrino with different mass scales can play important role in many cosmological phenomena. Recent results from MicroBooNE suggests the non-possibility for existence of an eV scale sterile neutrino. But, keV to GeV scale sterile neutrinos are still well-motivated theoretically and do not contradict any existing experiments\cite{2021microboone}.

\begin{table*}
\begin{center}
\renewcommand{\arraystretch}{1.5}
\begin{tabular}{c|c|c}
 \hline 
parameter &	best fit $\pm$ 1$\sigma$ &	3$\sigma$ range \\
\hline
$\vert\Delta m^2_{21}\vert: [10^{-5} eV^2]$ &	$7.41^{+0.21}_{-0.20}$  &	6.82–8.03 \\
$\vert\Delta m^2_{31}\vert: [10^{-3} eV^2] (NO)$	& $2.511^{+0.028}_{-0.027}$  & 2.428–2.597 \\
$\vert\Delta m^2_{32}\vert: [10^{-3} eV^2] (IO)$ &	$2.498^{+0.032}_{-0.025}$  & 2.408–2.581\\
$\sin^2\theta_{12} $	& $0.303^{+0.012}_{-0.011}$ & 0.270–0.341 \\
$\sin^2\theta_{23} (NO)$ &	$0.572^{+0.018}_{-0.023}$  &	0.406–0.620 \\

$\sin^2\theta_{23} (IO)$ &	$0.578^{+0.016}_{-0.021}$ 	&0.412–0.623 \\

$\sin^2\theta_{13} (NO)$ &	$0.02203^{+0.00056}_{-0.00059}$ 	&0.02029–0.02391 \\

$\sin^2\theta_{13} (IO)$ &	$0.02219^{+0.00060}_{-0.00057}$ 	&0.02047–0.02396 \\

$\delta_{\rm CP}/^o (NO)$ &	$197^{+42}_{-0.25}$	&	108-404 \\

$\delta_{\rm CP}/^o (IO)$ & $286^{+27}_{-32}$ &	192-360 \\
\hline
\end{tabular}  
\label{table1}
\end{center}
\caption{\centering Updated global-fit data for three neutrino oscillation, $Nufit$ 2022\cite{2020data}.}
\end{table*}

Cosmological and astrophysical measurements have indicated the presence of a mysterious, non-luminous, non-baryonic matter called dark matter(DM) which accounts for 26.8$\%$ of the total energy density of the Universe \cite{clowe2006direct,DM2017}. Yet, the fundamental nature of DM i,e. their origin, its constituents and interactions are still unknown. The requirements of a DM particle \cite{taoso2008dark} rule out all the SM particles. One of the most interesting beyond Standard Model(BSM) particle which behave as a warm dark matter (WDM) is the sterile neutrino. Particularly, sterile neutrinos with masses in the keV range having very small mixing with the active neutrinos of the order of $\sin^22\theta \sim \mathcal{O}(10^{-10})$ can be a dark matter \cite{Merle_2012,merle2017kev,dasgupta2021sterile}. According to the latest Planck data \cite{ade2016planck}, the relative abundance of DM in the Universe is observed as 
\begin{equation}
\Omega_{DM}h^2 = 0.1187 \pm 0.0017
\label{omegabound}
\end{equation} 
Dodelson-Widrow mechanism\cite{dodelson1994sterile} provides the correct abundance of keV-scale sterile neutrinos as DM. It requires only the mixing between active and heavier sterile neutrino, and thus this mechanism is very well-motivated. Further, the Shi-Fuller mechanism\cite{shi1999new} which includes lepton asymmetries, can result in resonantly enhanced production of sterile neutrinos. This allows it to access a larger range of the $m_s-\theta_s$ parameter space. The stability of keV sterile neutrino at the cosmological scale is one of the most important criteria to be a DM candidate. Sterile neutrinos can radiatively decay into an active neutrino and a monoenergetic photon through $\nu_s \rightarrow \nu_a + \gamma$. The decay rate is given by \cite{pal1982radiative,ng2019new}

\begin{equation}
\Gamma=1.38\times 10^{-32}\left(\frac{\sin^2 2\theta}{10^{-10}}\right) \left(\frac{m_s}{keV}\right)^5 s^{-1}
\label{gammaeq}
\end{equation}
where $\sin^2 2 \theta = 4\sum_{i}^{e,\mu,\tau} \vert U_{i4}\vert^2$ is the effective active-sterile mixing angle and $\vert U_{i4}\vert $ are the elements of active-sterile mixing matrix. The sterile neutrino lifetime is thus sufficiently large and it is essentially stable over timescales comparable to the age of the Universe $t_{U} = 4.4 \times 10^{17}$sec \cite{ade2016planck}, so that it is a good DM candidate. Following Ref.\cite{Das:2019kmn,abada2014dark}, the sterile neutrino abundance is expressed as a function of active-sterile mixing angle $\theta$ and the mass of sterile neutrino $m_s$ as 
\begin{equation}
\Omega_{DM}h^2 \simeq 0.3 \left(\frac{\sin^2 2\theta}{10^{-10}}\right) \left(\frac{m_s}{100keV}\right)^2
\label{omegaeq}
\end{equation}
There are constraints on the mass and mixing of the keV-scale sterile neutrino with active neutrinos provided by various cosmological and laboratory experiments such as observed DM relic density\cite{2021planck}, X-ray searches for sterile neutrino decay\cite{schneider2016astrophysical,boyarsky2006constraints,PhysRevD.75.063511,bulbul2014detection,10.1093/pasj/psv081}, Lyman-$\alpha$ forests\cite{baur2016lyman}, Tremaine-Gunn\cite{tremaine1979dynamical} and phase-space analysis\cite{boyarsky2009lower}, etc. Combining these constraints, we can infer that the mass of the sterile neutrino should be $m_s \geq 4$ keV, and its mixing with SM neutrinos should be $\sin^2 2\theta \leq 10^{-6}$\cite{arguelles2019production}. Various works on neutrino models based on the study of keV-mass sterile neutrinos as DM can be found in Refs.\cite{Berbig:2022nre,Boyarsky,Gautam1,Das2019,abada2014dark,Jaramillo}. Detailed analysis on the production and decay of keV-scale sterile neutrino DM in the early Universe are given in Refs.\cite{adhikari2017white,abazajian2017sterile,kusenko2009sterile,Boyarsky,DeGouvea:2019wpf}. Searches for decaying DM signal in the keV-MeV mass range have been conducted using a wide range of X-ray telescope, XMM-Newton\cite{boyarsky2006,boyarsky2008,boyarsky2016}, Chandra\cite{watson2012constraining,riemer2016constraints,hofmann20167}, Suzaku\cite{suzaku1,suzaku2},Swift\cite{swift}, NuStar\cite{neronov2016decaying,Riemer,perez2017almost} etc. There also has been an unknown observation of a 3.5 keV line in the X-ray spectra of galaxy clusters\cite{galaxycluster} as well as Andromeda\cite{andromeda} and Milky Way galaxies\cite{milkyway}. This could be interpreted as a signal coming from the decay of DM particle having mass $\sim 7$ keV\cite{MERLE2015283}. However, this result is widely disputed and future progress with our understanding of the origin and nature of this signal may come with the next generation of high-resolution X-ray missions, including XARM (Hitomi replacement mission), LYNX and Athena+\cite{neronov2016toward}. 

Another important problem that we shall address here is the origin of matter-antimatter asymmetry of the Universe. The current data on this Baryon Asymmetry of Universe (BAU) is given by 

\begin{equation}
Y_B = \frac{\eta_{B}-\eta_{\overline{B}}}{s} \simeq (8.7 \pm 0.06) \times 10^{-11}
\end{equation}
Fukugida and Yanagida\cite{asaka2002non,buchmuller2005leptogenesis} showed that if heavy right-handed Majorana neutrinos exist, they can decay in a lepton number violating process to the SM leptons and Higgs boson. If C and CP are violated in these decays, there can be symmetry between leptons and anti-leptons. Finally, the lepton asymmetry is converted into baryon asymmetry through sphaleron process and thus producing matter-antimatter asymmetry. Many authors have discussed baryogenesis via leptogenesis in different neutrino models\cite{Kalita:2015jaa,Das2019,Gautam2,goswami,Das3}.

Further, neutrinoless double beta decay $(0\nu\beta\beta)$  is also one of the most important experiments to probe the Dirac or Majorana nature of neutrinos\cite{abada2019beta,zeldo1981}. These experiments are also sensitive to the absolute neutrino masses through the effective neutrino mass $m_{\beta\beta}$. KAMLand-Zen\cite{Kamland} and GERDA provided an upper bound on $m_{\beta\beta}$ through a combined analysis in the range $m_{\beta\beta} < (0.071-0.161$) eV \cite{agostini2018improved}. There are kinematic measurements of $\beta-$decay also and the absolute effective neutrino mass is directly probed from the cut-off of the electron energy spectrum emitted from $\beta-$decay. Recent result from KATRIN experiment \cite{Katrin2020} constrains the effective electron neutrino mass $m_{\beta}$ to be less than 1.1 eV and the latest Planck data \cite{2021planck} provides the upper limit on the sum of active neutrino masses $\sum m_i < 0.12$eV. The parameters $m_{\beta\beta}$ and  $m_{\beta}$ are expressed as the sum of mass eigenstates and elements of lepton mixing matrix as \cite{Hagstot}.

\begin{equation}
m_{\beta\beta}=\vert\sum_{j=1}^4\vert U_{ej}\vert ^2m_j\vert,
\label{mbbeq}
\end{equation}

\begin{equation}
m_{\beta} = \left(\sum_{i=1}^4\vert U_{ei}\vert ^2 m_{i}^2\right)^{1/2}.
\label{mbeq}
\end{equation}

From these, we are motivated to study a neutrino mass model based on  minimal extended seesaw mechanism(MES) to explain some of the SM problems mentioned above. Several other literatures are available based on similar studies \cite{vien2021,vien20224,vien2022b,vien2022lepton,Gautam1,Gautam2,Gautam2020bnx,Das2019,das2019active} but the main shortcomings of such models are the presence of many hypothetical flavon fields as well as additional discrete symmetries. Particularly, in Ref.\cite{Das2019}, the authors considered different sets of five flavons  extended with another two flavons as perturbations in the Lagrangian seperately for normal heirarchy(NH) and inverted heirarchy(IH). In Ref.\cite{Gautam1}, authors use $S_4$ symmetry as an extension to SM along with $Z_4\times Z_3$ in Inverse Seesaw(ISS) scenario. They consider six flavons : four $S_4$ triplets and two singlets, in order to generate the mass matrices in the model.  The present work is different and more efficient in the fact that we use only four flavon fields $\phi, \psi, \zeta $ and $\chi$ to construct the Dirac, Majorana and sterile neutrino mass matrices. Deviation from $\mu-\tau$ symmetry is generated through another $A_4$ triplet flavon $\eta$. The fields are given group charges different from other works and hence a new structure of light neutrino mass matrix is obtained using the minimal extended seesaw formula. We also consider the mass of sterile neutrino in a broader range of $(4-50)$keV.

The detailed description of the model is given in the next section. The structure of this paper is as follows. We present a detailed description of the model along with the mechanism for generating neutrino masses in section \ref{section2} followed by the numerical analysis of the model in section \ref{section3}. Results of the analysis is presented in section \ref{results}. We conclude with a brief summary and discussion in section \ref{discussion}. 

\section{Description of the model}\label{section2}
In this model we have considered an extension of SM through $A_4\times Z_4 \times Z_2 $ where an $A_4$ singlet sterile neutrino $S$ is added along with three right handed neutrino singlets $\nu_{R1},\nu_{R2}, \nu_{R3}$. $A_4$ has four irreducible representations  denoted by singlets $1, 1'', 1'$ and a triplet $3$.  The SM lepton doublet $l$ transforms as triplet under $A_4$ while the charged lepton singlets $e_R, \mu_R, \tau_R$ transform as $1, 1'', 1'$ respectively. Two flavons $\phi$ and $\psi$ are used  along with three Higgs $H,H^{\prime}, H^{\prime\prime}$ to give neutrino masses through electroweak symmetry breaking. Another flavon $\eta$ is responsible for breaking the $\mu-\tau$ symmetry of neutrino mass matrix and to generate non-zero $\theta_{13}$. Two $A_4$ singlets $\chi$ and $\zeta$ are used to generate a diagonal Majorana mass matrix $M_R$ and sterile neutrino mass matrix $M_S$ respectively.  The particle contents and their corresponding group charges are shown in Table \ref{table2}. Additional discrete symmetry $Z_2$ is used in order to remove some unwanted interactions  in the Lagrangian which are otherwise allowed by $A_4\times Z_4,$ such as $\frac{1}{\Lambda}(y_m\overline{l}\tilde{H^{\prime}}\psi)_{1}\nu_{R1}$, $\frac{1}{\Lambda}(y_n\overline{l}\tilde{H^{\prime\prime}}\psi)_{1}S$, etc. The charged lepton and Dirac neutrino mass matrices are constructed using the Weinberg dim-5 operator\cite{Weinberg}.

The invariant Yukawa interaction terms of the model in the charged lepton sector are given by 
\begin{equation}
-\mathcal{L}^{cl}_{y} \sim \ \frac{y_e}{\Lambda}(\overline{l}H\psi)_1e_R +\frac{y_{\mu}}{\Lambda}(\overline{l}H\psi)_{1^{\prime}}\mu_R + \frac{y_{\tau}}{\Lambda}(\overline{l}H\psi)_{1^{\prime\prime}}\tau_R 
\end{equation}
Whereas, the Yukawa interaction for the neutrino sector which are invariant under all the symmetry groups of the model  are given by

\begin{align}\label{ns}
-\mathcal{L}^{mass}_{y} \sim & \ \frac{1}{\Lambda}(y_1\overline{l}\tilde{H}\phi+y_3\overline{l}\tilde{H}\eta)_{1}\nu_{R1} +\frac{1}{\Lambda}(y_1\overline{l}\tilde{H}^{\prime}\phi+y_3\overline{l}\tilde{H}^{\prime}\eta)_{1}\nu_{R2} \\ \nonumber &+\frac{1}{\Lambda}(y_4\overline{l}\tilde{H}\psi+y_2\overline{l}\tilde{H}^{\prime\prime}\phi + y_3\overline{l}\tilde{H}^{\prime\prime}\eta)_{1}\nu_{R3}\\ \nonumber
&+\ \frac{1}{2}\lambda_1\chi\overline{\nu}_{R1}^c\nu_{R1} + \frac{1}{2}\lambda_2\chi\overline{\nu}_{R2}^c\nu_{R2} +\frac{1}{2}\lambda_3\chi\overline{\nu}_{R3}^c\nu_{R3}\\ \nonumber
&+\ \frac{1}{2}\lambda_s\zeta\overline{S}^c\nu_{R1}
\end{align}
where $\tilde{H}=i\sigma_2H^*$ is used to make the Lagrangian Gauge invariant.

\begin{table*}[t]
\centering
\begin{tabular}{ccccccccccccccc}
\hline
$\frac{Fields}{Charges}$ & $l$ & $e_R, \mu_R,\tau_R$ & H & H$^{\prime}$ &H$^{\prime\prime}$ & $\psi$ &$\phi$ &$\chi$& $\zeta$ & $\nu_{R1}$&$ \ \nu_{R2} $&$\ \nu_{R3}$& S & $\eta$ \\
\hline
$SU(2)_L$& 2 &1  & 2 & 2 & 2 &1 &1 &1 &1 &1 &1 &1 &1&1  \\
$U(1)_{Y}$&-1/2 &+1  & -1/2&-1/2 & -1/2 &1 &1 &0 &-1 &0 &0 &0 &1&1  \\
$A_4$& 3 &1,1$^{\prime\prime}$,1$^{\prime}$ &1 &1$^\prime$ &1$^{\prime\prime}$ &3 &3 &1 &$1^{\prime\prime}$ &1 &1 &1 &1$^{\prime}$ &3 \\
$Z_4$&1 &1  &1 &-i & i &1 &-i &1 &-1 &i &-1 &1 &-i&-i \\
$Z_2$&1 &1  &1 &-1 & 1 &1 &1 &1 &-1 &1 &-1 &1 &-1&1 \\
\hline
\end{tabular}
\caption{\footnotesize Particle content of the model and their group charges.}
\label{table2}
\end{table*}

After electroweak symmetry breaking, the scalar flavon fields obtain their vacuum expectation values (v.e.v.) along with alignments given as \cite{pramanick2018three,KING2007351,Das3} 
\begin{align}
\langle\psi\rangle=v(1,0,0);\ \langle\eta\rangle=v(0,1,0); 
 \langle\phi\rangle=v(1,1,1);\ \langle\chi\rangle =v ;\ \langle\zeta\rangle=u
 \label{vev}
\end{align}

Using the multiplication rule of $A_4$\cite{ishimori2010non}, the charged lepton mass matrix becomes diagonal, 
\begin{equation}
M_L = \frac{\langle H\rangle v}{\Lambda}diag(y_e,y_{\mu},y_{\tau}).
\label{cl}
\end{equation}
As a result, the charged-lepton diagonalization matrix becomes unity, $U_L =1$. Thus, the lepton mixing matrix depends on the neutrino sector only.

From Eq.(\ref{ns}) using Eq.(\ref{vev}), the Dirac, Majorana and sterile neutrino mass matrices will take the form
\begin{eqnarray}
M_D=\left(
\begin{array}{ccc}
 a & a & c+h \\
 a & a & h \\
 a+t & a+t & h+t \\
\end{array}
\right),
\hspace{0.1cm} 
 M_R=\left(
\begin{array}{ccc}
 d & 0 & 0 \\
 0 & e & 0 \\
 0 & 0 & f \\
\end{array}
\right),
\label{MD}
\end{eqnarray}
\begin{equation}
M_s=\left(
\begin{array}{ccc}
 g & 0 & 0 \\
\end{array}
\right).
\label{Ms}
\end{equation}

where, 
$$a = \frac{\langle H\rangle v}{\Lambda}y_1,\ h = \frac{\langle H\rangle v}{\Lambda}y_2,\ c = \frac{\langle H\rangle v}{\Lambda}y_4,\ t = \frac{\langle H\rangle v}{\Lambda}y_3, \ 
 d =\lambda_1 v, $$ $$ e =\lambda_2 v,\ \ \ \ f =\lambda_3 v,\ \ \ g=\lambda_s u.$$

In order to achieve the sterile neutrino mass in the keV range, the v.e.v of scalar $\chi$ is assumed to lie around TeV scale. An approximate estimate of the mass scales of the parameters in the model are as follows, 
$$
\Lambda \sim 10^{15}\ \mbox{GeV}, \ \ v \sim 10^{14}\  \mbox{GeV} ,\ \ u \sim 10\ \mbox{TeV}\ \mbox{and}\ \langle H \rangle \sim 125\ \mbox{GeV}
$$

In our analysis, we have used minimal extended seesaw(MES) mechanism to calculate the masses of active neutrino as well as sterile neutrino. In MES, the $4\times 4$ active-sterile neutrino mass matrix is given by \cite{Zhang2011} 
\begin{equation}
M_{\nu}^{4\times 4} = -\left(\begin{matrix}
M_DM_R^{-1}M_D^T & M_DM_R^{-1}M_S^T \\
 
M_S(M_R^{-1})^TM_D^T & M_SM_R^{-1}M_S^T
\end{matrix} \right).
\label{4by4}
\end{equation}
It is important to observe that $det( M_{\nu}) =0$. Thus, at least one of the neutrino mass eigenvalues is zero in MES mechanism. Applying the seesaw condition, $M_{\nu}$ is further diagonalised and the $(3\times 3)$ active neutrino mass matrix $m_{\nu}$ and the sterile neutrino mass $m_s$ are expressed as
\begin{align}
m_{\nu} \simeq &\  M_DM_R^{-1}M_S^T\left(M_S M_R^{-1}M_S^T\right)^{-1}M_S\left(M_R^{-1}\right)^T M_D^T -M_DM_R^{-1}M_D^T\ ;
\label{m} \\ 
m_s \simeq & - M_SM_R^{-1}M_S^T.
\label{ms}
\end{align}
Using Eq.(\ref{m}) and Eq.(\ref{ms}), the active neutrino mass matrix and the sterile neutrino mass are given by 

\begin{equation}
m_{\nu}= -\left(
\begin{array}{ccc}
 \frac{a^2}{e}+\frac{(c+h)^2}{f} & \frac{a^2}{e}+\frac{h (c+h)}{f} & \frac{a (a+t)}{e}+\frac{(c+h) (h+t)}{f} \\
 \frac{a^2}{e}+\frac{h (c+h)}{f} & \frac{a^2}{e}+\frac{h^2}{f} & \frac{a (a+t)}{e}+\frac{h (h+t)}{f} \\
 \frac{a (a+t)}{e}+\frac{(c+h) (h+t)}{f} & \frac{a (a+t)}{e}+\frac{h (h+t)}{f} & \frac{(a+t)^2}{e}+\frac{(h+t)^2}{f} \\
\end{array}
\right)
\label{mv}
\end{equation}
and
\begin{equation}
m_s = -\frac{g^2}{d}.
\end{equation}

The $\mu-\tau $ symmetry in $m_{\nu}$ can be realised if we put $t=0$ in  eq.(\ref{mv}), i,e. 

\begin{equation}
m_{\nu}= -\left(
\begin{array}{ccc}
 \frac{a^2}{e}+\frac{(c+h)^2}{f} & \frac{a^2}{e}+\frac{h (c+h)}{f} & \frac{a^2}{e}+\frac{h (c+h)}{f} \\
 \frac{a^2}{e}+\frac{h (c+h)}{f} & \frac{a^2}{e}+\frac{h^2}{f} & \frac{a^2}{e}+\frac{h^2}{f} \\
 \frac{a^2}{e}+\frac{h (c+h)}{f} & \frac{a^2}{e}+\frac{h^2}{f} & \frac{a^2}{e}+\frac{h^2}{f} \\
\end{array}
\right)
\end{equation}

\section{Numerical analysis}\label{section3}
For numerical analysis, we use the latest 3$\sigma$ bounds of neutrino oscillation data shown in Table 3 for both normal hierarchy(NH) and inverted hierarchy(IH). Diagonal elements of the heavy Majorana mass matrix $M_R$ are given non-degenerate values $d=10^{13}$GeV , $e=10^{11}$GeV and $f=5\times 10^{11}$GeV.  We numerically diagonalise the active neutrino mass matrix $m_{\nu}$ using the relation $U^{\dagger}\mathcal{M}U = diag(m_1^2, m_2^2, m_3^2)$, where $\mathcal{M}=m_{\nu}m_{\nu}^{\dagger}$ and $U$ is a unitary mixing matrix. For  NH: $
m_1=0,\  m_2 =\sqrt{\Delta m_{21}^2},\mbox{ and}\
m_3=\sqrt{\Delta m_{21}^2 +\Delta m_{31}^2}$ and for IH: 
$
m_1=\sqrt{\Delta m_{21}^2 +\Delta m_{31}^2},\  m_2 =\sqrt{\Delta m_{21}^2},\
\mbox{and }m_3=0,
$ where $\Delta m_{ij}^2 = \vert m_j^2-m_i^2\vert.$ We can also define a parameter $r$ which is given by the ratio between the mass squared differences as 
\begin{equation}
r=\sqrt{\frac{\Delta m_{21}^2} {\Delta m_{31}^2}}= \frac{m_2}{m_3} \ \ \ \mbox{for\ NH}
\end{equation}
and 
\begin{equation}
r=\sqrt{\frac{\Delta m_{21}^2 }{\vert\Delta m_{32}^2\vert}}= \sqrt{1-\frac{m_1^2}{m_2^2}} \ \ \ \mbox{for\ IH}
\end{equation}

In PDG convention \cite{pdg2022}, $U$ is parameterised using three mixing angles $\theta_{12},\theta_{13},\theta_{23}$, one Dirac phase $\delta_{CP}$ and two Majorana phases $\alpha,\beta$. The values of $\alpha$ and $\beta$ are unknown and they are randomly  varied in the range (0, 2$\pi$). The light neutrino mass matrix $m_{\nu}$ contains four unknown complex parameters $a,c,t$ and $h$. We randomly choose the values of these parameters in the ranges given in Table \ref{table3}. The parameter space is constrained by upper bound on the sum of active neutrino mass $\sum m_i < 0.12$ eV and the 3$\sigma$ bounds of the neutrino mixing angles. The ratio $r$ is independent of the mass scales and mixing matrix $U$. From Table \ref{table1}, the best-fit value of $r$ in NH is $r_{o} = 0.172.$ The allowed parameter space is further constrained by the experimental values of $r$.

The full $(4\times 4)$ active-sterile mass matrix is diagonalised by a unitary $(4\times 4)$ mixing matrix given by \cite{v441982}
\begin{equation}
V \simeq \left(\begin{matrix}
(1-\frac{1}{2}RR^{\dagger})U & R \\ 
-R^{\dagger}U & 1-\frac{1}{2}R^{\dagger}R
\end{matrix} \right),
\label{V44}
\end{equation}
where $R$ represents the strength of active-sterile mixing given by
\begin{align}
R=&M_DM_R^{-1}M_S^T(M_SM_R^{-1}M_S^T)^{-1}\ =\ \left(
\begin{array}{c}
 \frac{a}{g} \\
 \frac{a}{g} \\
 \frac{a+t}{g} \\
\end{array}
\right)
\end{align}

We can solve the neutrino mixing angles from the elements of active-sterile mixing matrix $V$ \cite{DEV2019401,mksingh}. However, The deviation from unitarity of the $3\times 3$ mixing matrix $U$ is given by $\frac{1}{2}\vert  RR^{\dagger}\vert$ and it is found to be very small $\leq \mathcal{O}(10^{-10})$. As a result, we can ignore the effects of the keV-scale sterile neutrino in the mixing matrix $U$. Neutrino mixing angles are obtained from $U$ using the general formula given below  
\begin{align}
\sin^2\theta_{13}=\vert U_{13}\vert^2,\ \sin^2\theta_{12}=\frac{\vert U_{12}\vert^2}{1-\vert U_{13}\vert^2} , \ \sin^2\theta_{23}=\frac{\vert U_{23}\vert^2}{1-\vert U_{13}\vert^2}
\end{align}

In order to study the possibility of keV-scale sterile neutrino as a dark matter, it is required that the active-sterile mixing angle is very small, $\theta_s < 10^{-6}$. Remaining parameter $g$ is solved by constraining the sterile neutrino mass in the range $(4-50)$ keV.

Besides, one of the most important parameters in neutrino sector is the Jarlskog invariant $J$. Dirac CP violating phase $\delta_{CP}$ is related to $J$ and it is given by

\begin{equation}
J=Im[U_{e1}U_{\mu 2}U^*_{e2}U^*_{\mu 1}]= s_{23}c_{23}s_{12}c_{12}s_{13}c_{13}^2\sin\delta_{CP}
\label{j}
\end{equation}

where $s_{ij}=\sin\theta_{ij}$ and $c_{ij}=\cos\theta_{ij}$ are the neutrino mixing angles calculated from the model. Similarly, the Majorana phases are also evaluated from $U$ using the invariants $I_1$ and $I_2$ defined as follows 
\begin{align}
I_1 = Im[U_{e1}^*U_{e2}]=c_{12}s_{12}c^2_{13}\sin(\alpha/2), \\ 
I_2 = Im[U_{e1}^*U_{e3}]=c_{12}s_{12}c_{13}\sin\left(\frac{\beta}{2}-\delta_{CP}\right)
\end{align}

To find the best-fit values of the free parameters in our model as well as the neutrino observables, we use the $\chi^2$ function given by 
\begin{equation}
\chi^2(x_i) = \sum_{j}\left(\frac{y_j(x_i)-y_j^{bf}}{\sigma_j}\right)^2
\label{chitest}
\end{equation}
 where $x_i$ are the free parameters in the model and $j$ is summed over the observables $\{\sin^2\theta_{12},\sin^2\theta_{13},\sin^2\theta_{23},r\}$. $y_j(x_i)$ denotes the model predictions for the observables and $y_j^{bf}$ are their best-fit values obtained from the global analysis. $\sigma_j$ denotes the corresponding uncertainties obtained by symmetrizing $1\sigma$ range of the neutrino observables given in table \ref{table1}. By minimizing the overall $\chi^2$ function, we can calculate the best-fit values of our model parameters.
 
\begin{table*}
\centering
\begin{tabular}{c|cc|cc}
\hline 
Parameters $\ \ $ & Allowed ranges (GeV) $\ $ & Best-fit (GeV) & Parameter phases & Best-fit \\ 
\hline 
$\vert a\vert$ & 0.46 - 0.60 & 0.574 &$\phi_a = (0, 2\pi)$  & 0.59$\pi$  \\ 
$\vert c\vert$ & 2.53 - 4.72 & 3.835  &  $\phi_c = (0, 2\pi)$ & 0.44$\pi$\\ 
$\vert t\vert$ & 0.80 - 1.59 & 0.949 &  $\phi_t = (0, 2\pi)$& 2.19$\pi$\\ 
$\vert h\vert$ & 3.20 - 3.94 &3.469  & $\phi_h = (0, 2\pi)$& -2.98$\pi$ \\  
\hline
\end{tabular} 
\caption{\footnotesize Allowed ranges of model parameters and their best-fit values corresponding to $\chi^2_{min}$.}
\label{table3}
\end{table*} 
 
\begin{figure}
\subfigure[]{
\includegraphics[width=.45\textwidth]{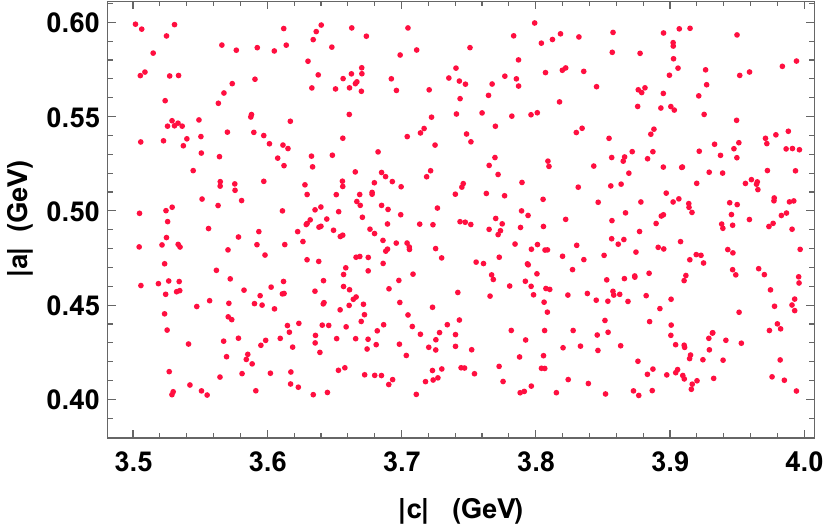}}
\quad
\subfigure[]{
\includegraphics[width=.45\textwidth]{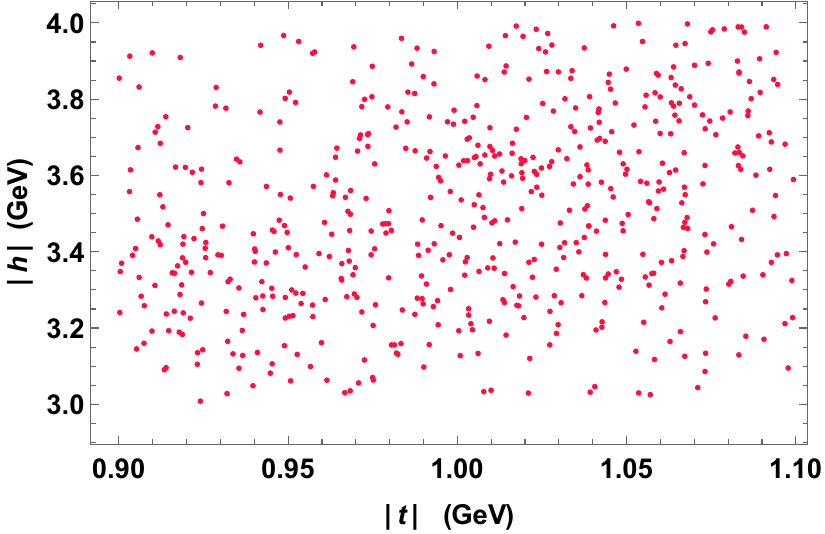}}
\quad
\subfigure[]{
\includegraphics[width=.45\textwidth]{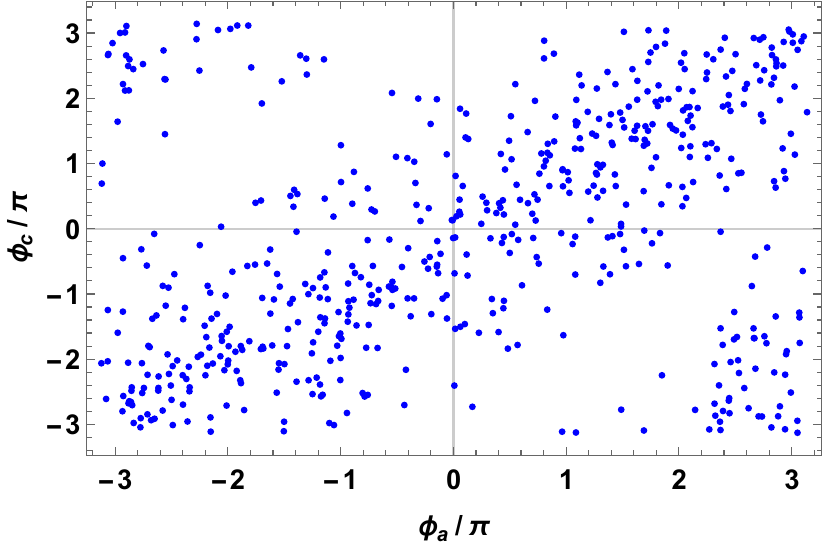}}
\quad
\subfigure[]{
\includegraphics[width=.45\textwidth]{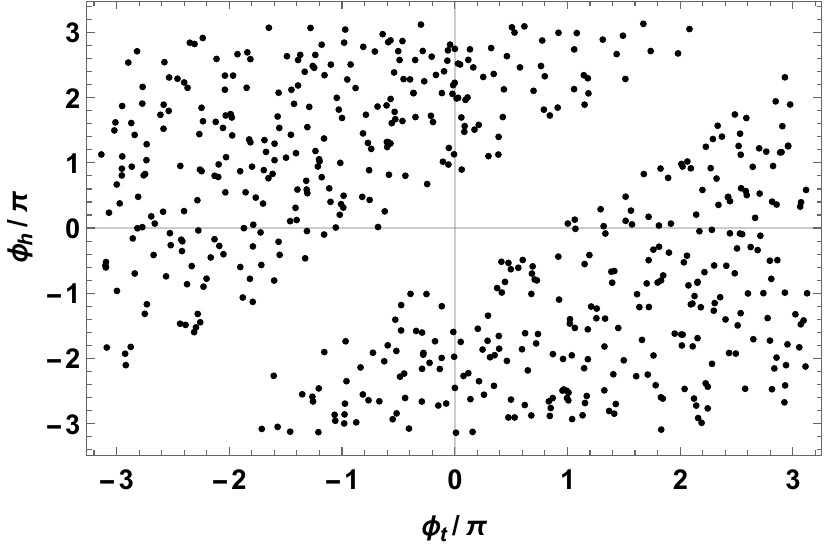}}
\quad
\caption{\footnotesize{(a),(b). Variation among allowed model parameters. (c),(d). Variation between the allowed phases of the parameters.}}
\label{parameters}
\end{figure}

\begin{figure}
\subfigure[]{
\includegraphics[width=.45\textwidth]{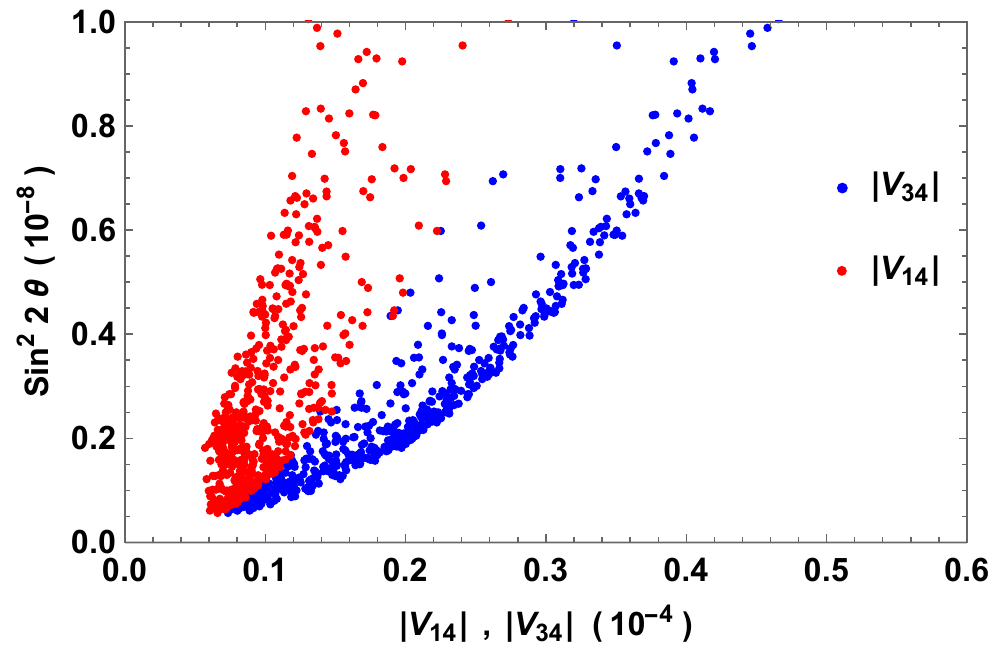}}
\quad
\subfigure[]{
\includegraphics[width=.45\textwidth]{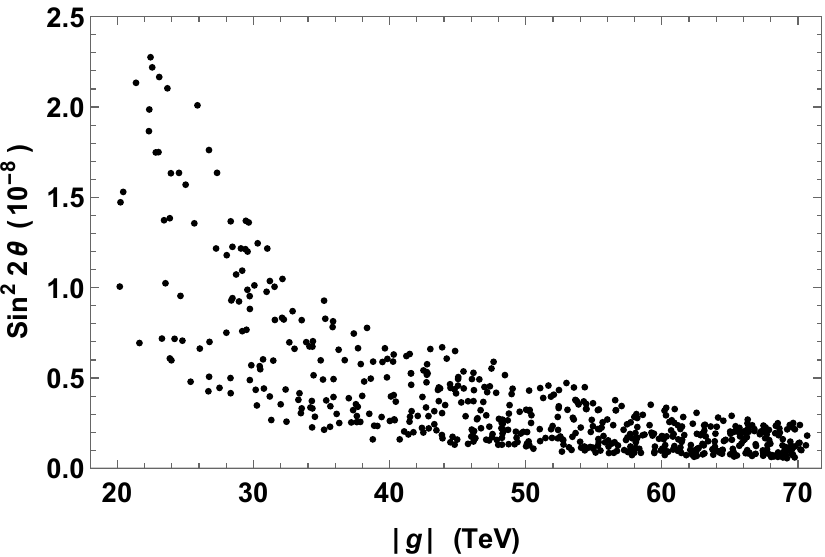}}
\quad
\caption{\footnotesize{(a). Variations of effective mixing angle $\sin^2 2 \theta$ with elements of active-sterile mixing strength $R$. (b) Variation of  $\sin^2 2 \theta$ with $|g|$.}}
\label{v14eff}
\end{figure}

\begin{figure}
\subfigure[]{
\includegraphics[width=.46\textwidth]{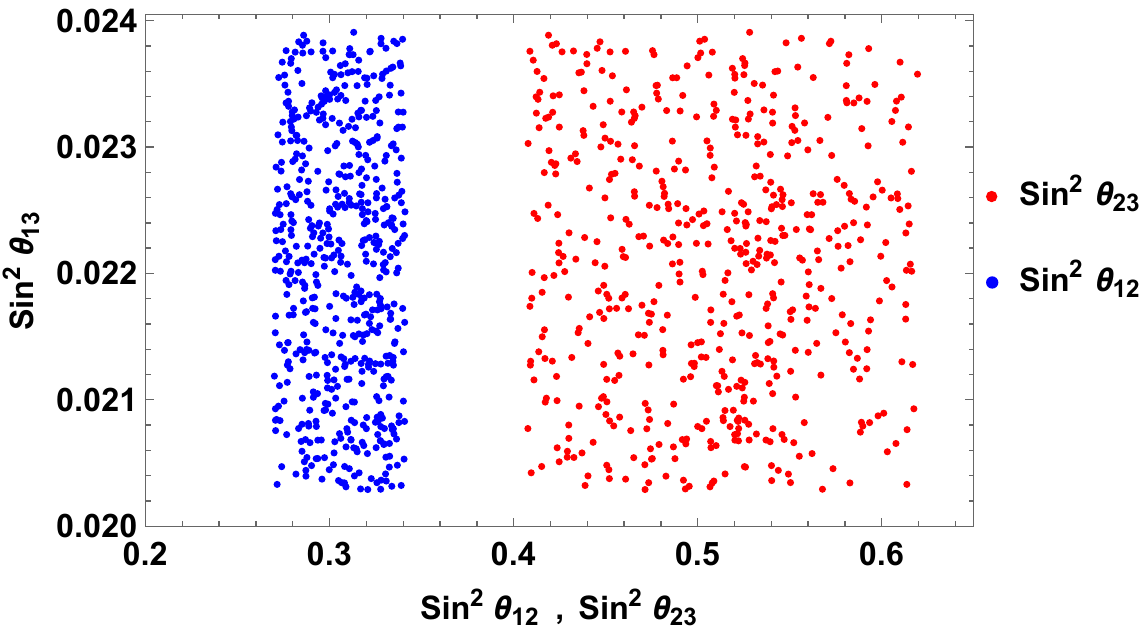}}
\quad
\subfigure[]{
\includegraphics[width=.42\textwidth]{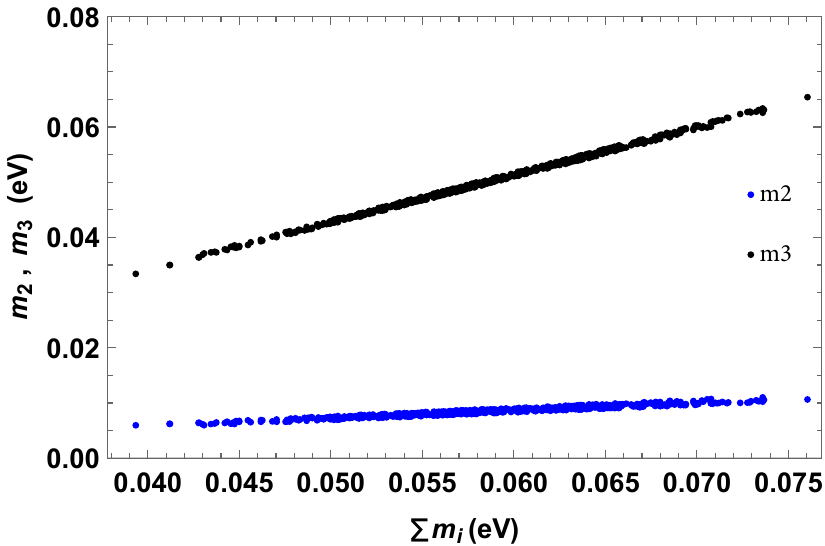}}
\quad
\caption{\footnotesize{(a). Plot between mixing angle $\sin^2 \theta_{13}$ with  $\sin^2 \theta_{12}$ and  $\sin^2 \theta_{23}$. (b) Variation of  $m_1$ and $m_2$ with $\sum m_i $.}}
\label{angleandmass}
\end{figure}

\begin{figure}
\subfigure[]{
\includegraphics[width=.45\textwidth]{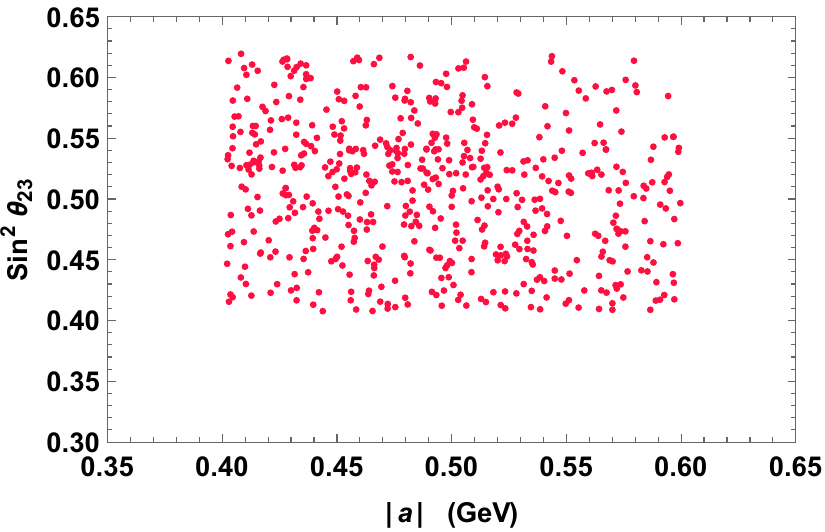}}
\quad
\subfigure[]{
\includegraphics[width=.45\textwidth]{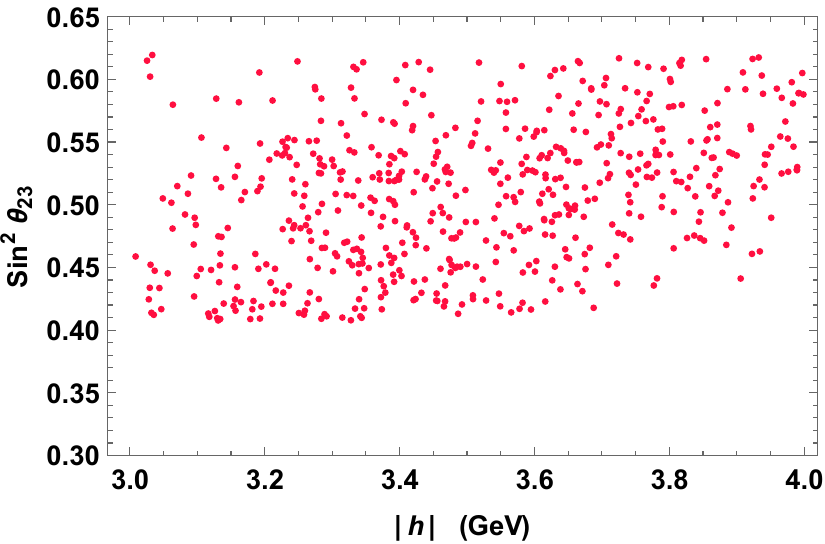}}
\quad
\caption{\footnotesize{ Plot between mixing angle $\sin^2 \theta_{13}$ with model parameters $\vert a\vert$ and $\vert h\vert.$}}
\label{s23ah}
\end{figure}

\section{Results}\label{results}
For the given ranges of the parameters in Table \ref{table3}, our model is compatible with global fit results of the neutrino oscillation parameters at 3$\sigma$ level for NH only, while IH is disallowed at $3\sigma$ level. The variation between model parameters allowed in the numerical analysis for NH are shown in Fig.\ref{parameters} (a),(b). The phases of the allowed parameters are shown as scatter plots in Fig.\ref{parameters} (c),(d). It is observed that the allowed parameter space is very narrow for $a$ and $t$ compared to $c$ and $h$. Fig.\ref{angleandmass} (b) shows the plot between active neutrino masses $m_2$ and $m_3$  with the sum of neutrino masses $\sum m_i$. For NH, $m_1 =0$ and upper bound on the sum of neutrino mass is obtained at $\sum m_i \leq 0.07604$ eV while the maximum values of $m_2$ and $m_3$ are found to be respectively $0.01103$ eV and $0.06541$ eV. From the $\chi^2$ analysis, we find that the best fit values of the model parameters for $\chi^2_{min}=0.8575$ in NH are obtained at $\vert a \vert = 5.74 \times 10^{8}\mbox{eV} ,\vert c \vert = 3.83 \times 10^{9} \mbox{eV},\vert t \vert = 9.49\times 10^{8}\mbox{eV}\ \mbox{and}\ \vert h \vert = 3.47\times 10^{9} \mbox{eV} $ while the respective best-fit values of their phases are obtained at $\phi_a = 0.591 \pi ,\phi_c = 0.443 \pi , \phi_t = 2.19\pi$ and $\phi_h =-2.98\pi $. Corresponding best-fit values of neutrino observables are found to be $\sin^2\theta_{12} = 0.30110, \sin^2\theta_{13}=0.02212, \sin^2\theta_{23}=0.55542$ and $r = 0.17466 .$  The complete parameter ranges and their best-fit values determined from the numerical analysis of the model are shown in Table \ref{table3}. Effective active-sterile mixing angle $\sin^22\theta$ as a function of parameter $|g|$ and mixing matrix element $\vert V_{14}\vert, \vert V_{34}\vert $ are shown in Fig.\ref{v14eff}.

For the charged lepton sector, by comparing Eq.(\ref{cl}) with the experimental values for
masses of the charged leptons given in Ref.\cite{pdg2022}, $m_e =
0.51099$ MeV, $m_{\mu} = 105.65837$ MeV, $m_{\tau}= 1776.86$ MeV,
we get the charged lepton Yukawa constants as $y_e \sim 10^{-5}, \ y_{\mu}\sim 10^{-3}$ and $y_{\tau} \sim 10^{-2}$.

\begin{figure}
\subfigure[]{
\includegraphics[width=.45\textwidth]{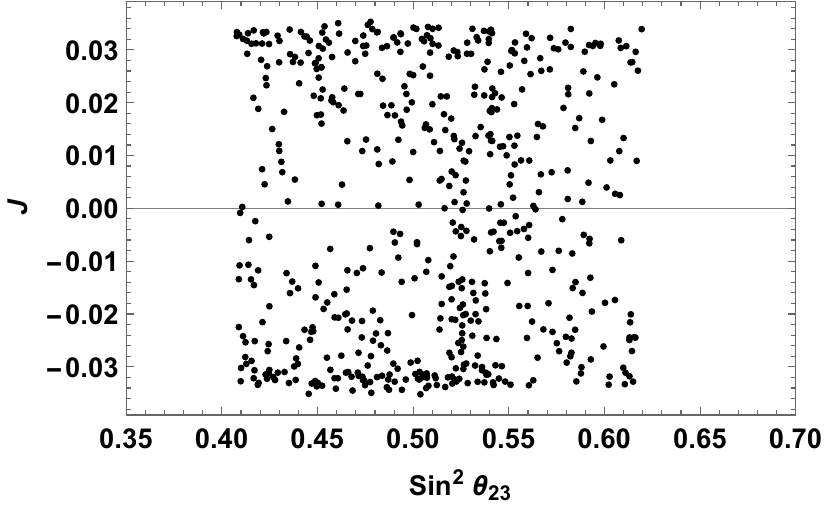}}
\quad
\subfigure[]{
\includegraphics[width=.45\textwidth]{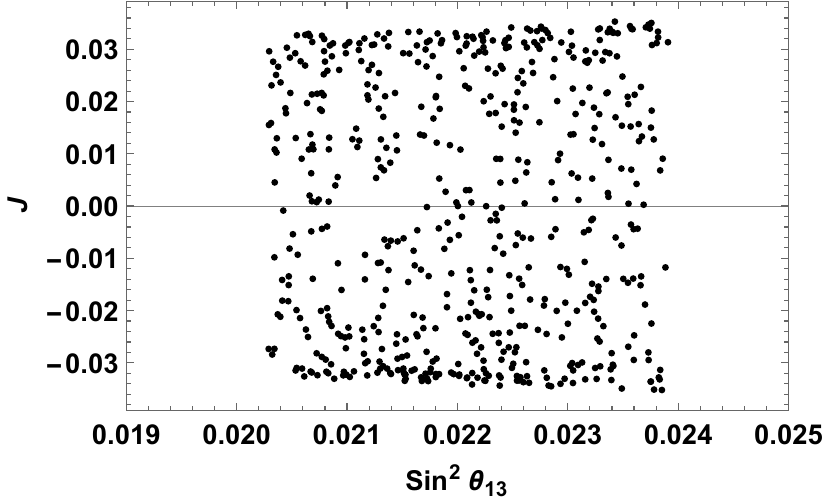}}
\quad
\subfigure[]{
\includegraphics[width=.45\textwidth]{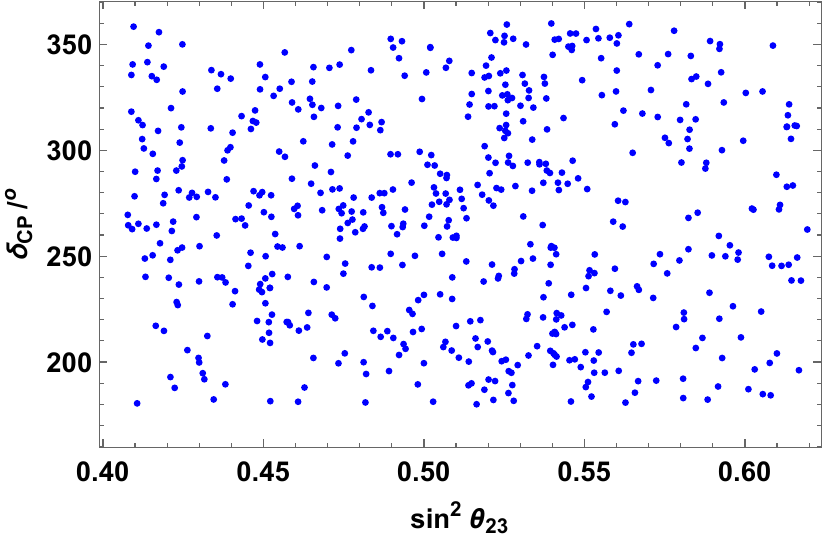}}
\quad
\subfigure[]{
\includegraphics[width=.45\textwidth]{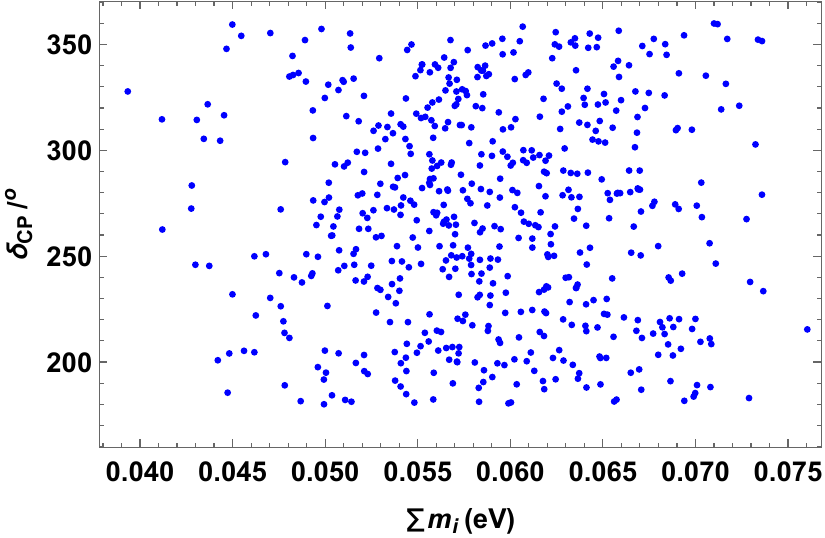}}
\quad
\caption{\footnotesize{(a),(b). Plot between mixing angles $\sin^2 \theta_{13}$ and $\sin^2 \theta_{23}$ with $J$. (c),(d). Variation of CP-violating Dirac phase $\delta_{CP}$ with $\sin^2\theta_{23}$ and $\sum m_{i}$.}}
\label{delta}
\end{figure}

\begin{figure}
\subfigure[]{
\includegraphics[width=.47\textwidth]{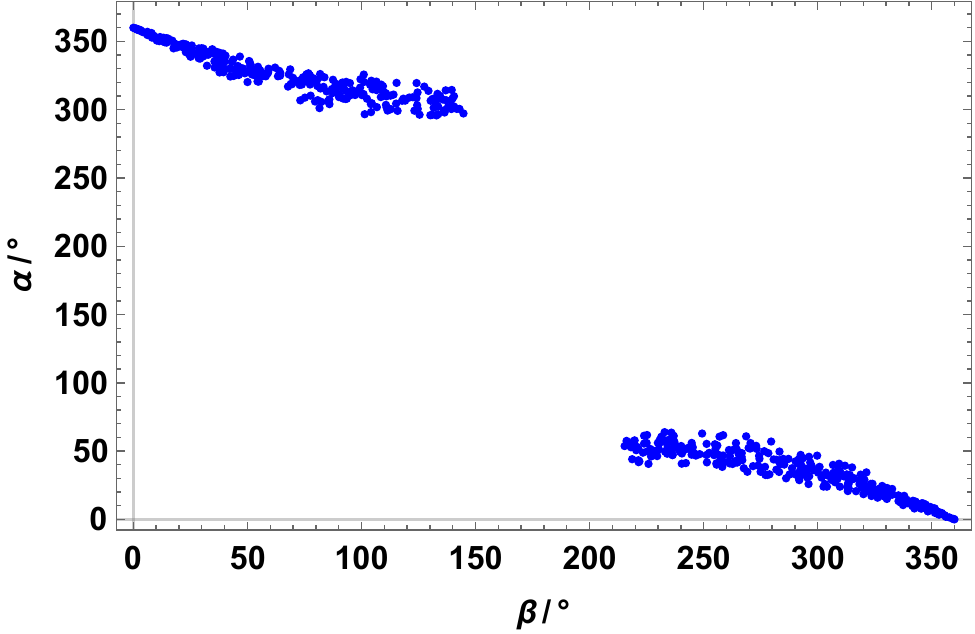}}
\quad
\subfigure[]{
\includegraphics[width=.40\textwidth]{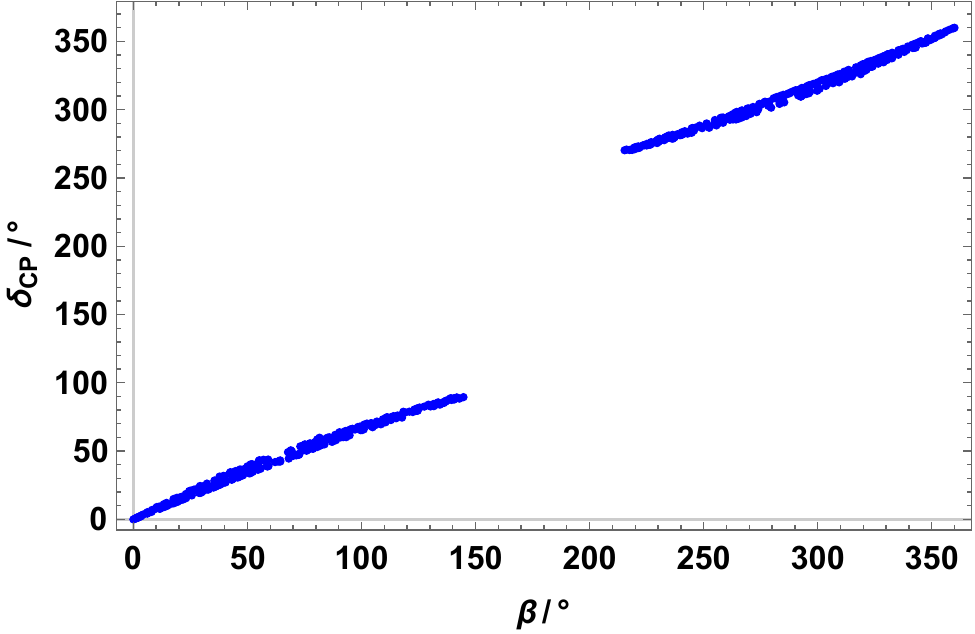}}
\quad
\caption{\footnotesize{(a). Plot between the Majorana phases $\alpha$ and  $\beta$. (b) Variation of  $\delta_{CP}$  with Majorana phase $\beta $.}}
\label{alphabeta}
\end{figure}

\begin{figure}
\subfigure[]{
\includegraphics[width=.45\textwidth]{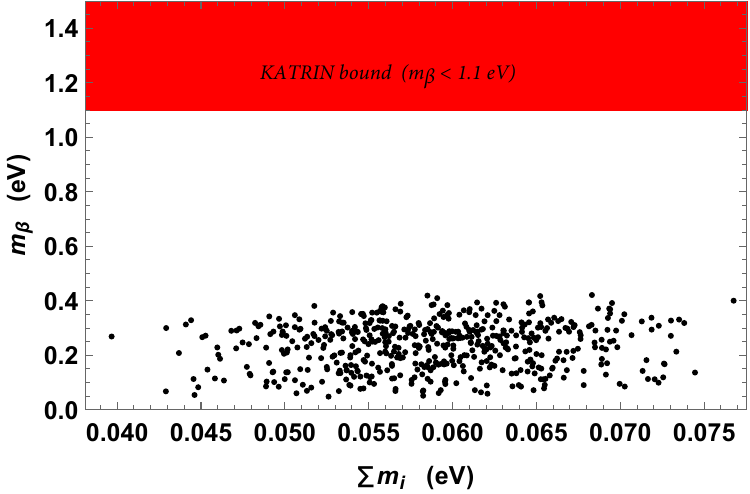}}
\quad
\subfigure[]{
\includegraphics[width=.45\textwidth]{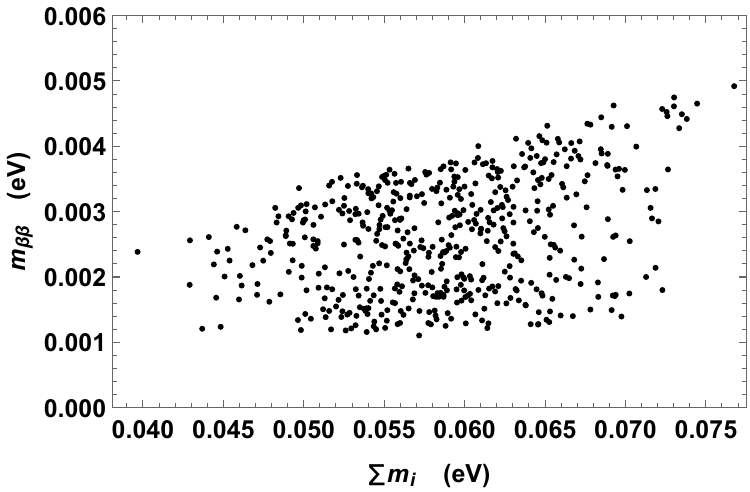}}
\quad
\caption{\footnotesize{(a). Plot between effective electron mass $m_{\beta}$ with $\sum m_{i}$. The coloured strip shows the upper bound from latest KATRIN data $m_{\beta} < 1.1$eV.  (b) Variation of effective neutrino mass $m_{\beta\beta}$ with $\sum m_i $.}}
\label{mbb}
\end{figure}

\subsection{Mixing angles and phases $\delta_{CP}$, $\alpha $ and $\beta $}
The variation plot among neutrino mixing angles $\sin^2 \theta_{12}$, $\sin^2 \theta_{23}$ with  $\sin^2 \theta_{13}$  is shown in Fig.\ref{angleandmass}(a). Values of  $\sin^2 \theta_{23}$ are distributed within the 3$\sigma$ range with very small crowding towards region above 0.5. Future generation neutrino oscillation experiments such as JUNO \cite{JUNO},Hyper-Kamiokande \cite{abe2018hyper}, DUNE \cite{abi2020deep}, etc. will hopefully confirm the neutrino mass ordering and also determine the octant degeneracy of $\theta_{23}$. However, data points for  $\sin^2 \theta_{13}$ and  $\sin^2 \theta_{12}$ are equally distributed within their $3\sigma$ ranges. In order to show the variation of neutrino mixing angles as a function of the model parameters, we have plotted $\sin^2 \theta_{23}$ with the model parameters $\vert a\vert$ and $\vert h\vert$ in Fig.\ref{s23ah}. It can be observed that larger value of $\vert h\vert$ results in larger values of $\sin^2 \theta_{23}$ whereas values of $\sin^2 \theta_{23}$ does not show any distinct correlation with parameter $a$.

Dirac CP-violating phase $\delta_{CP}$ is calculated through the Jarlskog invariant $J$ as given in Eq. (\ref{j}). We have plotted $J$ as a function of $\sin^2 \theta_{23}$ and $\sin^2 \theta_{13}$ in Fig.\ref{delta} (a),(b). Values of $J$ determined from the model lie within the range $\pm 0.0352$. Fig.\ref{delta} (c),(d) show the plot between $\delta_{CP}$ with $\sin^2\theta_{23}$ and $\sum m_{i}$ respectively. We obtain the Dirac CP-violating phase in the range $180.01^{o}\leq \delta_{CP}\leq 359.96^{o}$. The best fit-value of $\delta_{CP}$ is obtained at $326.09^{o}$.  Fig.\ref{alphabeta} shows the predictions of Majorana phases $\alpha$ and $\beta$.

\subsection{Effective masses $m_{\beta}$ and $m_{\beta\beta}$ from $\beta$-decay and  $0\nu\beta\beta$}
The effective neutrino mass and effective electron mass are calculated from $0\nu\beta\beta$ and $\beta$-decay experiments using Eq.(\ref{mbbeq}) and Eq.(\ref{mbeq}) respectively. Variations of $m_{\beta\beta}$ and $m_{\beta}$ with sum of active neutrino masses $\sum m_i$ are shown in Fig.\ref{mbb}. From these plots, we observe that the effective mass parameter lies in the range $m_{\beta\beta} \sim (0.9794-5.0291)$ meV  which will be a great challenge for future $0\nu\beta\beta$ experiments \cite{dolinski2019neutrinoless,cao2020towards}. On the other hand, Fig\ref{mbb}.(a) indicates that the effective electron mass is obtained in the range $m_{\beta} \sim (0.0845-0.4103)$ eV which well below the upper bound provided by the latest KATRIN experiment \cite{Katrin2020}.

\subsection{Dark matter}
\begin{figure}
\subfigure[]{
\includegraphics[width=.47\textwidth]{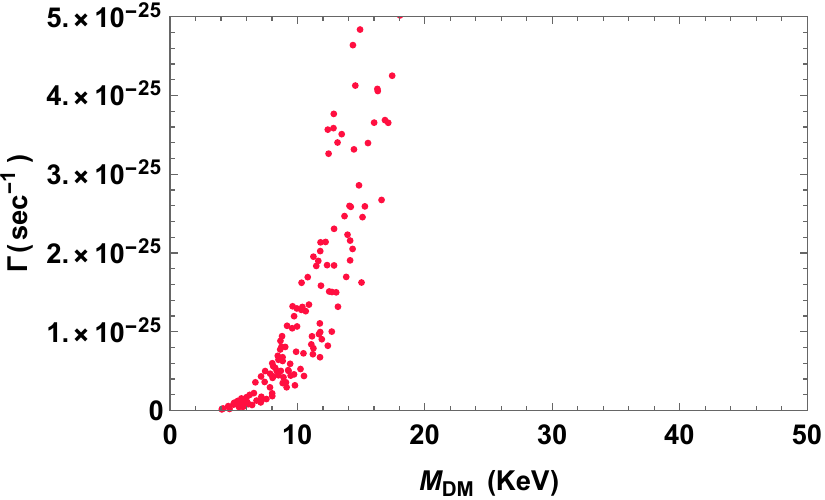}}
\quad
\subfigure[]{
\includegraphics[width=.435\textwidth]{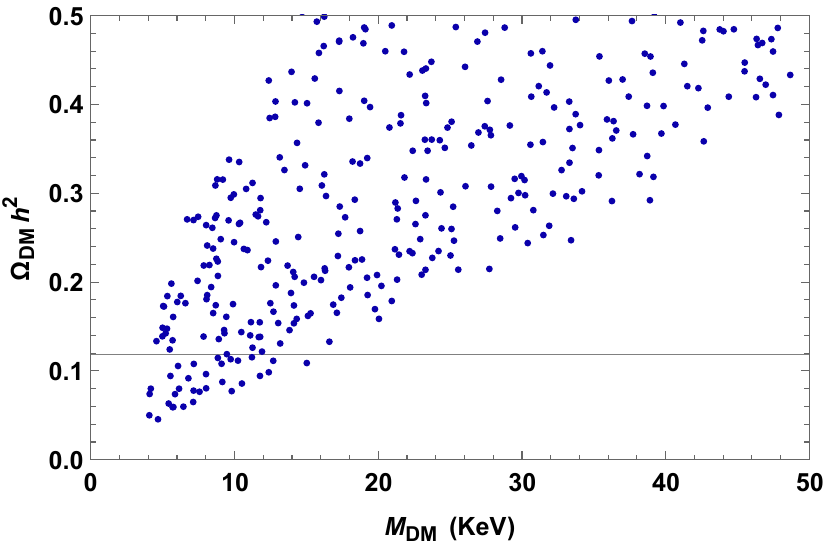}}
\quad
\caption{\footnotesize{(a) Variation of decay width $\Gamma$ with mass of the sterile neutrino $m_s \equiv M_{DM}$. (b) Variation of relic density $\Omega_{DM}h^2$ with mass of sterile neutrino. Here, the horizontal line represents the experimental data of relic abundance of DM in the Universe which is $0.1187\pm 0.0017$}}
\label{dm1}
\end{figure}

\begin{figure}
\subfigure[]{
\includegraphics[width=.45\textwidth]{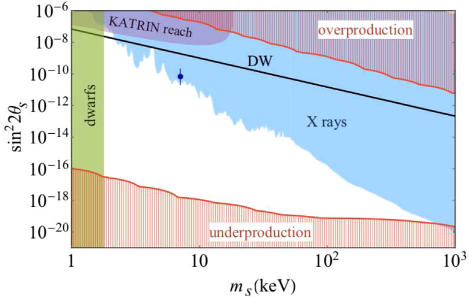}}
\quad
\subfigure[]{
\includegraphics[width=.45\textwidth]{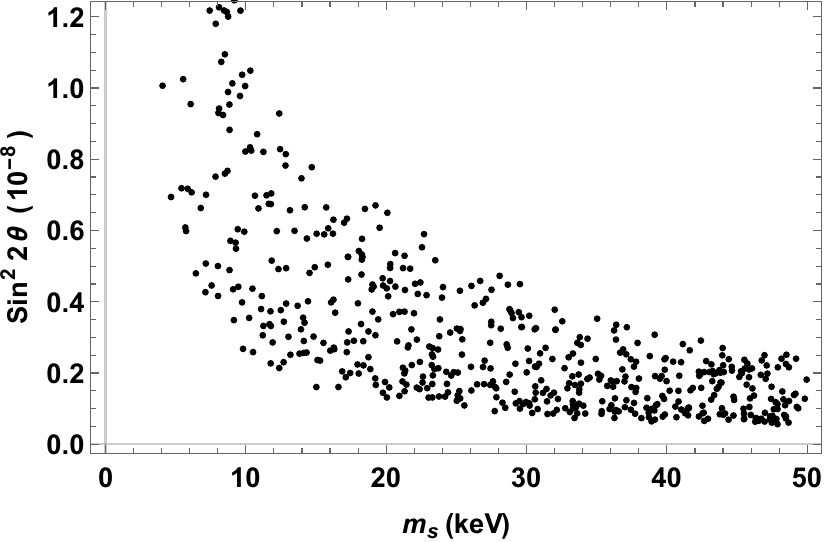}}
\quad
\caption{\footnotesize{(a).Various bounds on the mass and effective mixing angle of the keV sterile neutrino DM. This plot is taken as reference from Ref.\cite{de2020dodelson}. (b) Variation of effective mixing angle $\sin^22\theta$ with mass of sterile neutrino $m_s$.}}
\label{dm2}
\end{figure}

Assuming the existence of keV-scale sterile neutrinos, we work out the possibilities of it behaving as a dark matter candidate. We have used the relation for relic density and decay width from Eq.(\ref{omegaeq}) and Eq.(\ref{gammaeq}) respectively to check the validity of the present model in giving the observed DM density. Fig.\ref{dm1} (a) shows the plot between decay width $\Gamma$ with the mass of sterile neutrino $m_s \equiv M_{DM}$. In our analysis, we have considered the upper limit of decay width to be $\mathcal{O}(10^{-25}) sec^{-1}$ for the sterile neutrino to behave as a dark matter. From Fig.\ref{dm1} (a), it is found that the allowed mass range for sterile neutrino is $m_s \leq 20$ keV. This range of $m_s$ corresponds to the effective mixing angle $\sim \mathcal{O}(10^{-11}-10^{-9})$. Comparing with the results of the analysis in Ref.\cite{de2020dodelson} shown in Fig.\ref{dm2}(a), we can observe that the narrow mass range $m_s \leq 20\ \mbox{keV}$ corresponding to effective mixing angle $\mathcal{O}\sim (10^{-10}-10^{-11})$ is still allowed by the various experimental bounds. This observation may also supplement the unknown observation of 3.5 keV signal in X-ray data from the decay of a 7 keV sterile neutrino\cite{bulbul2014detection,andromeda}.  Similarly, based on the plots of decay width $\Omega_{DM}h^2$  in Fig.\ref{dm1}(b), we find that the allowed mass of sterile neutrino in a more constrained range of $(4-18)$ keV is observed to be giving the correct abundance consistent with current experimental value given in Eq.(\ref{omegabound}).


\subsection{Baryogenesis via Leptogenesis}
We have considered non-degenerate masses of heavy right-handed neutrinos such that $e<f<<d$, i,e. $\nu_{R2}$ is the lightest in our case. The lightest $\nu_{R2}$ can decay into a Higgs and lepton pair which will produce sufficient lepton asymmetry giving rise to the experimentally observed baryon asymmetry of the Universe. We have used the parameterization from Ref.  \cite{davidson2008leptogenesis} and the baryon asymmetry is given by
\begin{equation}
Y_B = c k \frac{\epsilon_{22}}{g*}
\end{equation}
where $c\sim 12/37$ is a constant that determines  the fraction of lepton asymmetry converted to baryon asymmetry. $k$ is the dilution factor which can be parameterized in our model as 
\begin{equation}
k\simeq\ \frac{0.3}{K(ln K)^{0.6}}
\end{equation}
where $K$ is defined as 

\begin{equation}
K=\frac{\Gamma}{H(T=M_2)}=\frac{(\lambda^{\dagger}\lambda)_{22}M_2}{8\pi}\frac{M_{Planck}}{1.66\sqrt{g*}M_2^2}
\end{equation}

\begin{figure}
\subfigure[]{
\includegraphics[width=.47\textwidth]{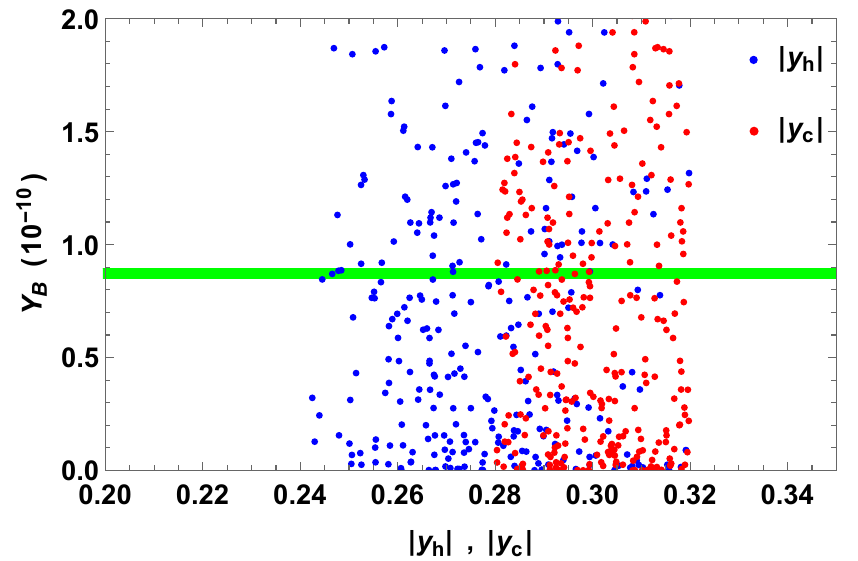}}
\quad
\subfigure[]{
\includegraphics[width=.435\textwidth]{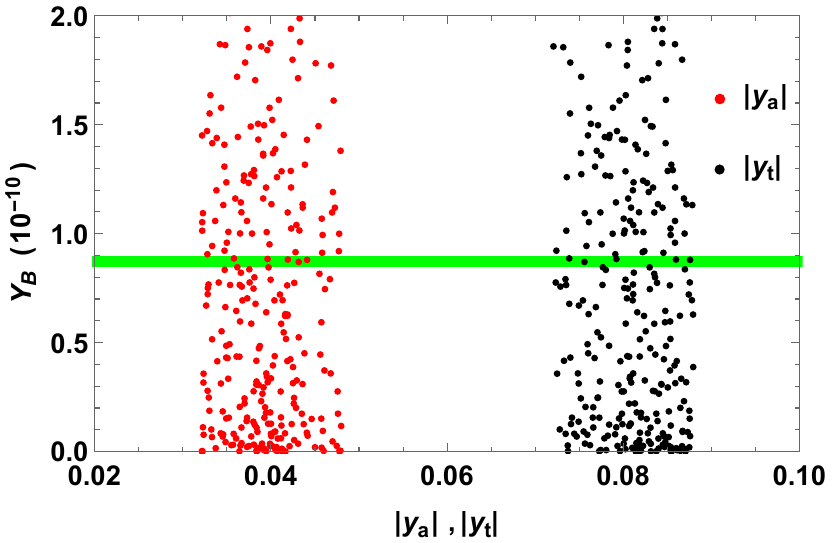}}
\quad
\caption{\footnotesize{Plot between Baryon asymmetry parameter $Y_B$ with the Yukawa couplings present in the lepton sector of the model. The green strip represents the experimental bound of $Y_B$ which is $(8.7\pm0.06)\times 10^{-11}$.}}
\label{YBplot}
\end{figure}

Here, $\Gamma$ is the decay width of $\nu_{R2}$. Further, $g*$ is the massless relativistic degree of freedom in the thermal bath and it is approximately 110. Finally, $\epsilon_{22}$ is the lepton asymmetry produced in the decay of the lightest $\nu_{R2}$. It is given by 
\begin{equation}
\epsilon_{22}=\frac{\Gamma(\nu_{R2}\rightarrow l_L + \bar{\phi})-\Gamma(\nu_{R2}\rightarrow \bar{l_L} + \phi)}{\Gamma(\nu_{R2}\rightarrow l_L + \bar{\phi})+\Gamma(\nu_{R2}\rightarrow \bar{l_L} + \phi)} 
\end{equation} 
The lepton asymmetry is calculated as \cite{davidson2008leptogenesis,covi1996cp,nardi2006importance,Engelhard:2007kf}
\begin{equation}
\epsilon_{22} = \frac{1}{8\pi}\frac{1}{(\lambda^{\dagger}\lambda)_{22}}\sum_{j}^{1,3} Im[(\lambda^{\dagger}\lambda)_{2j}]^2g(x_j)
\end{equation}
where $x_j \equiv \frac{M_j^2}{M_2^2}$ and within the SM \cite{covi1996cp} $g(x_j)$ is defined as,

\begin{equation}
g(x_j)=\sqrt{x_j}\left(\frac{2-x_j-(1-x_j^2)ln((1+x_j)/x_j)}{1-x_j}\right)
\end{equation} 
In the above relations, $\lambda$ is the Yukawa matrix corresponding to the Dirac mass matrix Eq.(\ref{MD}). We constructed the Yukawa matrix $\lambda$ from the model parameters $a, c, h $ and $t$.   We have solved $Y_B$ using these relations and the variation plots of $Y_B$ with Yukawa constants are shown in Fig.\ref{YBplot}. In our analysis, the Yukawa constants of the lepton sector are observed to be $\mathcal{O}(10^{-2}-10^{-1})$ which are in the acceptable range of the perturbativity limit $y\leq \sqrt{4\pi}$. The model parameters and their respective Yukawa couplings are found to give observed experimental bound for Baryon asymmetry $Y_B$.

\section{Summary and discussion}\label{discussion}
We have successfully developed a neutrino mass model based on the extension of SM using $A_4\times Z_4\times Z_2$ symmetry. We use the Weinberg dimension five operator to construct the neutrino mass matrices in the model. Possibility of a keV scale sterile neutrino to behave as a dark matter is studied using MES mechanism. The mass of the sterile neutrino is constrained in the range $(4-50)$keV. We solved the model parameters using the 3$\sigma$ values of the neutrino observables. It is observed that the model parameters can produce an effective active-sterile mixing $\sin^2\theta \leq 10^{-6}$ which is a requirement for sterile neutrino to be DM. We are able to reproduce neutrino oscillation parameters within their 3$\sigma$ range including the Planck bound on the sum of active neutrinos $\sum m_i <0.12$eV. Other phenomenological studies such as effective neutrino mass and effective electron mass from $0\nu\beta\beta$ and $\beta-$decay experiments, Baryogenesis via Leptogenesis, etc. are carried out to check the validity of our model. The effective mass parameter $m_{\beta\beta}$ is observed to be in the range $\sim (0.9794-5.0291)$meV while effective neutrino mass is found to be $m_{\beta} \sim (0.0845-0.4103)$ eV which is less than the upper bound of latest KATRIN data. The keV sterile neutrino mass in range $m_s \sim (4-18)$keV is found to give the correct abundance as well as decay width for a DM particle. Baryogenesis via leptogenesis is studied from the lepton asymmetry produced by the decay of the lightest Majorana right-handed neutrino which is $\nu_{R2}$ in our case. The experimental value of Baryon asymmetry $Y_B$ is found to be consistent for a very small range of Yukawa couplings $y_a, y_c, y_t$ and $y_t$. We have also determined the best-fit values of the model parameters as well as the neutrino observables using the $\chi^2$ analysis.
Further, it is important to note that our analysis is found to be consistent with the global analysis of neutrino oscillation data for NH only  while IH is excluded at the 3$\sigma$ level. In fact, for the given parameter space that we considered, the neutrino observables such as $r,$ gives larger values outside the 3$\sigma$. Hence, IH case is disallowed in this model structure with the given parameter space. However, it may be checked by considering a different set of parameter spaces or by making a small change in the structure of the model through new flavons \cite{Zhang2011,das2019active,Das3}, which can be studied in future. A small part of this work with similar model structure has been published in our previous work as a part of $NuDM-2022$ conference\cite{nudm}, where we studied the possibility of keV DM in a narrower mass range of $(1-18.5)$keV only. However, the parameter $r$ was not imposed as a constraint in the previous work and IH could be reproduced. In this work, we have expanded our study with other phenomenological studies and stricter constraints. In conclusion, we have presented a model which provides a well-motivated DM candidate as well as consistent neutrino phenomenologies.

\section*{Acknowledgments}
 One of the authors (MKS), would like to thank DST-INSPIRE, Govt. of India, for providing financial support under DST-INSPIRE Fellowship (IF180349).
 
\bibliographystyle{unsrt}
\bibliography{kevdark}

\end{document}